\documentstyle[prd,aps,epsfig,eqsecnum,amssymb,amsmath,floats]{revtex}

\begin{document}
\title{Constraints on power spectrum of density fluctuations from PBH evaporations}

\author{
Edgar Bugaev and Peter Klimai }

\address{ \bigskip
Institute for Nuclear Research, Russian Academy of Sciences,
\\ 60th October Anniversary Prospect 7a, 117312 Moscow, Russia }


\maketitle
\begin{abstract}

We calculate neutrino and photon energy spectra in extragalactic
space from evaporation of primordial black holes, assuming that
the power spectrum of primordial density fluctuations has a strong
bump in the region of small scales. The constraints on the
parameters of this bump based on neutrino and photon cosmic
background data are obtained.

\end{abstract}

\section{Introduction}

It is well known that for a sufficient production of primordial
black holes (PBHs) in early Universe the spectrum of density
perturbations set down by inflation must be "blue", i.e., it must
have more power on small scales. This implies that the spectral
index of the scalar perturbations must be larger than 1, in strong
contradiction with the latest WMAP results \cite{WMAP}. In
particular, inflationary models of hybrid type, in which the
inflaton is trapped in a local minimum of the potential and which
predict blue spectra seem to be excluded as a possible source of
PBHs.

As an alternative one can consider the models in which the power
on small scales is boosted by means of a bump in the power
spectrum of primordial fluctuations, as suggested, e.g., in
\cite{Ivanov:1994pa}. Such a bump is predicted in many scenarios
of two-step inflation (see, e.g., %
\cite{Yamaguchi,Kawasaki}) and can, in principle, exist even in
single-field inflationary models (if the potential has a special
form leading to the period of fast-roll at the end of inflation
\cite{CE}.

In the present work we reconsider the problem of constraining the
power spectrum of primordial fluctuations calculating the process
of the formation of PBHs having small masses ($10^{11} -
10^{15}$g). Products of evaporation of these PBHs contribute, in
particular, to extragalactic photon and neutrino diffuse
backgrounds (which are measured experimentally).

\section{PBH mass spectrum}

Using the Press-Schechter formalism \cite{PS}, the differential
mass spectrum of
primordial black holes can be written in the form %
\cite{BugaevD55,BugaevD66}:

\begin{equation}
\label{nBH} n_{BH}(M_{BH}) = \sqrt{\frac{2}{\pi}} \rho_i \int
\frac{1}{M\sigma_H(M)} \Big( \frac{2}{3 M} - \frac{1}{\sigma_H(M)}
\frac {\partial \sigma_H(M)} {\partial M} \Big) %
\Big( \frac{(\delta^H_R)^2}{\sigma_H^2(M)} -1 \Big)\cdot %
e^{-\frac {(\delta^H_R)^2}{2\sigma_H^2(M)}} \cdot %
\frac{d\delta^H_R}{df(M, \delta^H_R)/dM}.
\end{equation}
One must note, before the explanation of the notations used in Eq.
(\ref{nBH}), that in the problem studied here it is convenient to
specify the spectrum of primordial fluctuations at a fixed time
rather than at horizon crossing. It is so because we assume that,
at the end of inflation, reheating and subsequent formation of the
perturbation spectrum occur almost instantaneously. The connection
between density contrasts at a fixed time and at horizon crossing
(neglecting, at the moment, a dependence of the gravitational
potential on the time) is quite simple \cite{Kolb},
\begin{equation}
\label{drhorhot} \Big( \frac {\delta\rho}{\rho} \Big)_t = %
\Big( \frac{M}{M_{hor}(t)} \Big) ^{-2/3} \cdot \Big( \frac
{\delta\rho}{\rho} \Big)_{hor}.
\end{equation}
Correspondingly, the mean square deviation (after smoothing the
density field on a given mass $M$) at horizon crossing,
$\sigma_H(M)$, is given by the formula
\begin{equation}
\label{sigmaH2M} %
\sigma_H^2(M) = \Big( \frac{M}{M_i} \Big) ^{4/3} \int \Big(
\frac{k}{a_i H_i} \Big) ^4 %
\delta_H^2(k) W^2(kR) T^2(k) \frac{dk}{k} \equiv %
\Big( \frac{M}{M_i} \Big) ^{4/3}  \sigma_R^2(M).
\end{equation}
Here, $M_i$ is the horizon mass at the end of inflation, $a_i$ and
$H_i$ are the scale factor and Hubble parameter at the end of
inflation, $R$ is the comoving smoothing scale, $R\equiv
1/k_{fl}$, connected with $M$ by the expression
\begin{equation}
\label{kflaiHi} %
\Big( \frac{M}{M_i} \Big) ^{-2/3} = \frac{k_{fl}^2}{(a_i H_i)^2}.
\end{equation}
The notations used for the other functions in the integral
(\ref{sigmaH2M}) are standard: $\delta_H(k)$ is the horizon
crossing amplitude for primordial perturbations of the density
contrast, $W(kR)$ and $T(k)$ are window and transfer functions.

The variable $\delta_R^H$ in Eq. (\ref{nBH}) is a value of the
(smoothed) density contrast at horizon crossing, $\rho_i$ is the
energy density at the end of inflation. At last, the function
$f(M, \delta_R^H)$ connects the values of the smoothing mass $M$,
density contrast $\delta_R^H$ and PBH mass $M_{BH}$,
\begin{equation}
\label{fMdelta} %
M_{BH} = f(M, \delta_R^H; M_i).
\end{equation}
The concrete form of this function depends on the features of the
gravitational collapse leading to PBH formation.

It is assumed that the dependence of $\delta_H(k)$ on $k$ at small
scales can be parameterized in the form
\begin{equation}
\label{dHparam} %
\lg \delta_H(k) = A + (\lg \delta_H^0 - A) \exp \Big[-\frac{(\lg k
- \lg k_0)^2}{2 \Sigma^2} \Big].
\end{equation}
Here, the value of $A$ is known from observations at large scales,
$\delta_H^0$, $k_0$ and $\Sigma$ are free parameters, the values
of which should be constrained.

\begin{figure}[!t]
\centerline{\epsfig{file=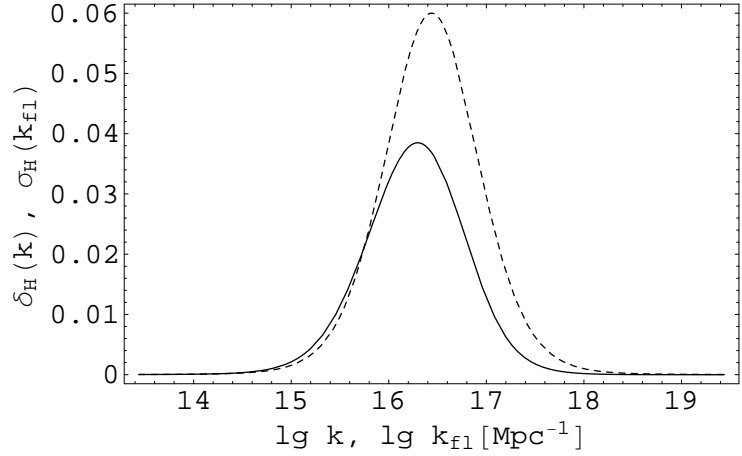, trim = 0 -12 0 0, width=0.55 \columnwidth} }%
\caption{Horizon crossing amplitude $\delta_H(k)$ (dashed line)
and smoothed dispersion $\sigma_H(k_{fl})$ (solid line). Spectra
are shown for $\Sigma=3, \delta_H^0= 0.06$, $k_0=2.75 \cdot
10^{16} Mpc^{-1}$.} \label{sigmafig}
\end{figure}

The calculation of $\sigma_H(M)$ was performed using the gaussian
window function,
\begin{equation}
\label{gauss} %
W(kR) = \exp \Big(-\frac{k^2 R^2}{2} \Big)
\end{equation}
and the transfer function, derived in the cosmological
perturbation theory,
\begin{equation}
\label{Tktau} %
T(k,\tau)= \frac{3}{(\omega_s \tau)^3} (\sin \omega_s\tau -
\omega_s\tau \cos \omega_s\tau ),
\end{equation}
\begin{equation}
\label{omegas} %
\omega_s = k c_s = \frac{k}{\sqrt{3}},
\end{equation}
$\tau$ is the conformal time. In the following we will ignore the
time dependence of the transfer function, putting
\begin{equation}
\label{Tktau2} %
T(k,\tau) \approx T\big(k, \tau=\frac{1}{k_0} \big) \equiv T(k)
\end{equation}
(because the most abundant PBH production takes place at the time
when the scale $k_0^{-1}$ corresponding to the maximum of
$\delta_H(k)$ crosses horizon).

In the model of spherically-symmetric collapse one obtains the
expression
\begin{equation}
\label{MBHsph} %
M_{BH} =  (\delta_R^H)^{1/2} M_h,
\end{equation}
where $M_h$ is the fluctuation (smoothing) mass at the moment of
horizon crossing. According to Carr and Hawking
\cite{CarrHawking}, $1/3 \le \delta_R^H \le 1$.

The connection between $M$ and the horizon mass $M_h$ is
\cite{BugaevD55}
\begin{equation}
\label{MhMiM} %
M_h = M_i^{1/3} M^{2/3}.
\end{equation}

From Eqs. (\ref{MBHsph}), (\ref{MhMiM}) one has the expression for
the function $f(M, \delta_R^H; M_i)$ for the Carr-Hawking
collapse:
\begin{equation}
\label{fCH} %
M_{BH} = f(M, \delta_R^H; M_i) = (\delta_R^H)^{1/2} M^{2/3}
M_i^{1/3}.
\end{equation}
In the picture of the critical collapse \cite{NJ} the
corresponding function is
\begin{equation}
\label{fcrit} %
f(M, \delta_R^H; M_i) = k_c (\delta_R^H-\delta_c)^{\gamma_c}
M^{2/3} M_i^{1/3},
\end{equation}
where $\delta_c$, $\gamma_c$ and $k_c$ are model parameters. In
this work we will accept the following set of parameters, which is
in agreement with recent calculations \cite{Musco}:
\begin{equation}
\label{parcrit} \delta_c=0.45, \;\;\;\; \gamma_c=0.36, \;\;\;\;
k_c=4.
\end{equation}

\begin{figure}
\centerline{\epsfig{file=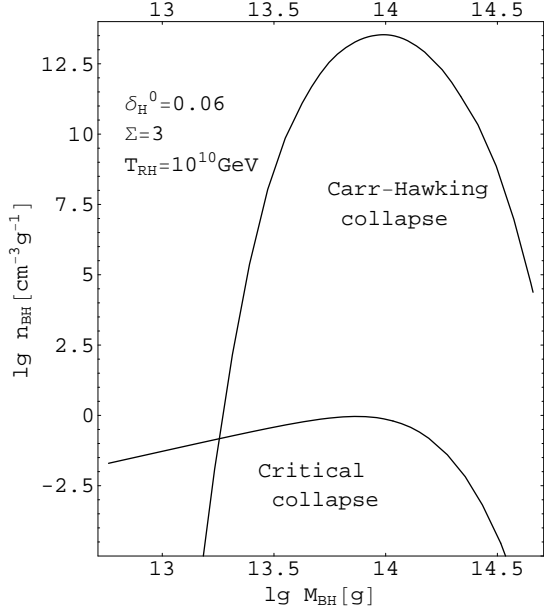, trim = 0 -12 0 0,  width=0.4 \columnwidth}} %
\caption{PBH mass spectra calculated for the models of critical
collapse and Carr-Hawking collapse. Here, we used the following
set of parameters: $\Sigma=3, \delta_H^0= 0.06, k_0=2.75 \cdot
10^{16} Mpc^{-1}, T_{RH}=10^{10} GeV.$ The parameters for the
critical collapse were taken as follows: $\delta_c=0.45,
\gamma_c=0.36, k_c=4$.} \label{nBHfig}
\end{figure}

Some results of PBH mass spectrum calculations are shown in Fig.
\ref{sigmafig} and \ref{nBHfig}. In Fig. \ref{sigmafig}, the
smoothed dispersion is drawn as a function of the inverse comoving
smoothing scale, $k_{fl}=1/R$. The connection between $k_{fl}$ and
horizon mass $M_h$ is determined from Eqs. (\ref{kflaiHi}) and
(\ref{MhMiM}) and is
\begin{equation}
\label{MhkflMiaiHi} M_h = \frac{M_i (a_i H_i)^2}{k_{fl}^2} .
\end{equation}

Fig. \ref{nBHfig} shows that the resulting PBH mass spectrum
strongly depends on the model of the gravitational collapse (at
the same values of the fluctuation spectrum parameters and
reheating temperature $T_{RH}$).

\begin{figure}
\centerline{ \epsfig{file=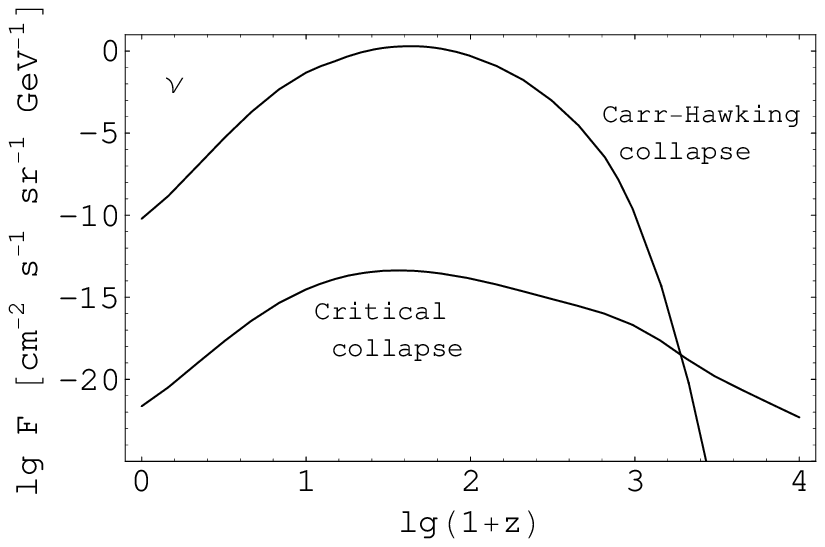, trim = 0 -12 0 0, width=0.45 \columnwidth} %
\epsfig{file=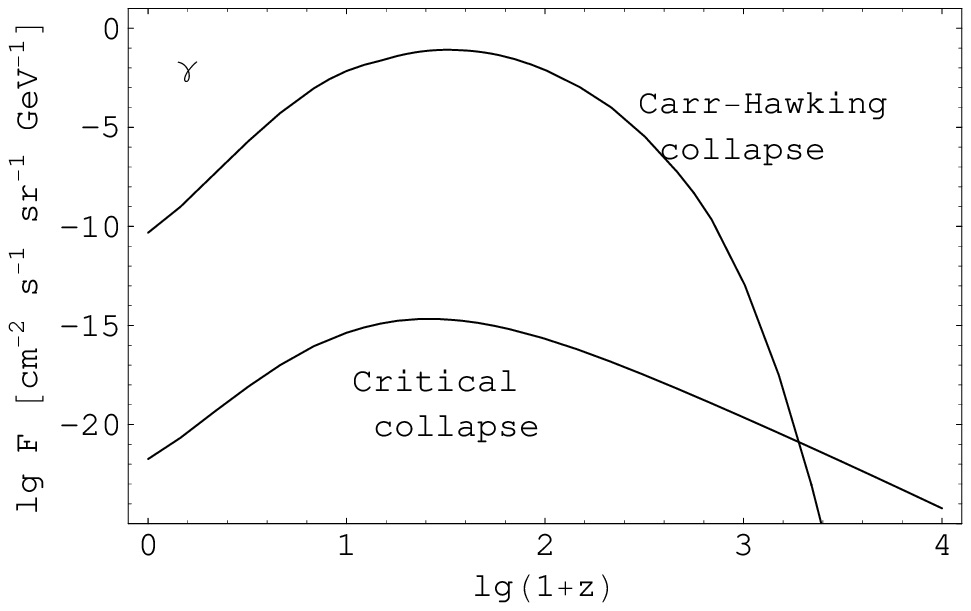, trim = 0 -12 0 0, width=0.41 \columnwidth} %
} \caption{Red shift distribution of the integrand $F(E,z)$. Left
panel is for neutrinos, right panel - for photons (absorption of
$\gamma$-rays at $z\ge 700$ is not shown in this figure). PBH mass
spectra shown in Fig. \ref{nBHfig} were used, $E=1GeV$ for both
panels. } \label{Ffig}
\end{figure}


\section{Neutrino and photon spectra from PBHs evaporations}

Evolution of a PBH mass spectrum due to the evaporation leads to
the approximate expression for this spectrum at any moment of
time:
\begin{equation}
\label{nBHmt} n_{BH}(m,t)=\frac{m^2}{(3\alpha t + m^3)^{2/3}}
n_{BH} \left((3\alpha t + m^3)^{1/3}\right),
\end{equation}
where $\alpha$ accounts for the degrees of freedom of evaporated
particles and, strictly speaking, is a function of a running value
of the PBH mass $m$. In all our numerical calculations we use the
approximation
\begin{equation}
\label{alphaconst} \alpha=const=\alpha (M_{BH}^{max}),
\end{equation}
where $M_{BH}^{max}$ is the value of $M_{BH}$ in the initial mass
spectrum corresponding to a maximum of this spectrum. Special
study shows that errors connected with such an approximation are
rather small.

The expression for an extragalactic differential energy spectrum
of neutrinos or photons (the total contribution of all black
holes) integrated over time is \cite{BugaevD55}
\begin{eqnarray}
\label{SE} S(E)=\frac{c}{4\pi}\int dt
\frac{a_0}{a}\left(\frac{a_i}{a_0}\right)^{3} \int dm
\frac{m^2}{(3\alpha t + m^3)^{2/3}} n_{BH} \left[(3\alpha t +
m^3)^{1/3}\right]
\cdot \varphi(E\cdot (1+z),m) e^{-\tau(E,z)} \equiv \nonumber\\
\\
\equiv \int F(E,z)d \lg (z+1).\nonumber
\end{eqnarray}

In this formula, $a_i$, $a$, and $a_0$ are cosmic scale factors at
$t_i$, $t$ and at present time, respectively, and $\varphi(E,m)$
is a total instantaneous spectrum of the radiation (neutrinos or
photons) from the evaporation of an individual black hole. The
exponential factor in Eq. (\ref{SE}) takes into account an
absorption of the radiation during its propagation in space. The
processes of neutrino absorption are considered, in a given
context, in \cite{BugaevD55}.
In the last line of Eq. (\ref{SE}) we changed the variable $t$ on
$z$ using the flat cosmological model with $\Omega_{\Lambda}\ne 0$
for which
\begin{eqnarray}
\label{dtdz} \frac{dt}{dz}=-\frac{1}{H_0 (1+z)}\left(\Omega_m
(1+z)^{3}+\Omega_r (1+z)^{4} + \Omega_\Lambda\right)
^{-1/2}, 
\end{eqnarray}
with $\Omega_r = (2.4\cdot 10^{4} h^2)^{-1}$, $h=0.7$,
$\Omega_M=0.3$, $\Omega_{\Lambda}=1-\Omega_M-\Omega_r$.

The instantaneous spectra of neutrinos and photons from PBH
evaporations were calculated using the photosphere model
elaborated in works \cite{Daghigh2001gy,Daghigh2002fn}. The
decoupling temperature $T_f$ for photons was taken equal to 120
MeV, and the temperature of the neutrinosphere, $T_{\nu}$, is
equal to 100 GeV, as in \cite{Daghigh2002fn}. The neutrino
emission from "cold" PBHs, those with $T_H < 200 $ GeV, was
calculated using the standard Hawking formula. The contribution to
the black hole neutrino spectra from decays of pions and muons
was, at this stage, not taken into account.

The photosphere model of black hole evaporation used here predicts
very steep time-integrated spectra of photons and neutrino
($E^{-4}$), and in the low energy region the absolute spectrum
values (the number of particles) are very high (comparing with the
models without photosphere). The average energy of photons and
neutrinos as a function of PBH lifetime is proportional to the
square root of black hole temperature ($\bar E \sim \sqrt{T_H}$
rather than $\bar E \sim T_H$).

On Fig. \ref{Ffig} the red shift distributions of the differential
energy spectra $S(E)$ are shown for the neutrino or photon energy
1 GeV. One can see that the distributions are more wide in the
case of the critical collapse (because the PBH mass spectrum in
this case has a long "tail" of small masses). The absorption
factor $e^{-\tau}$ is sufficient: for photons $z_{max} \sim 700$
\cite{Zdz} and for neutrinos it is about $10^6$ \cite{BugaevD55}.

\begin{figure}[!t]
\centerline{\epsfig{file=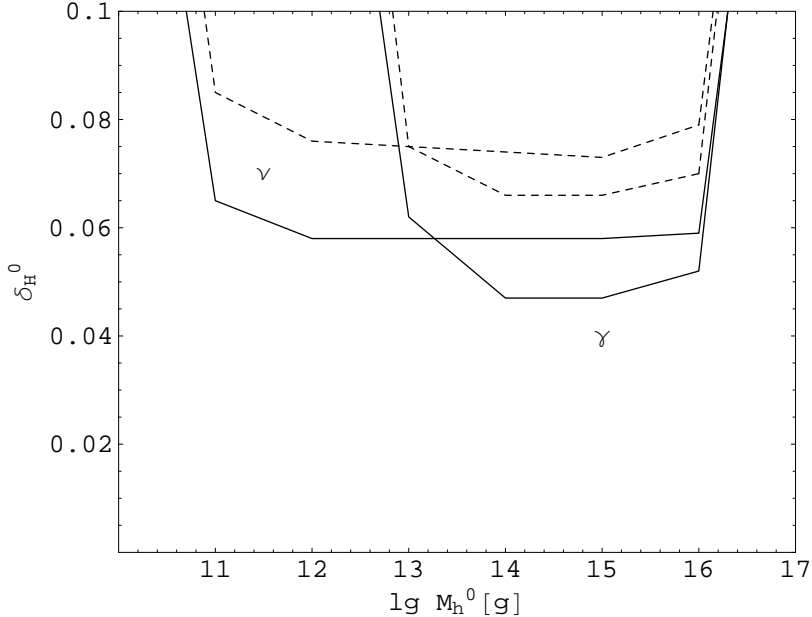, trim = 0 -12 0 0, width=0.6 \columnwidth}} %
\caption{Constraints on the horizon-crossing amplitude
$\delta_H(k)$ that were obtained in this work. $M_h^0$ is the
horizon mass at the moment when fluctuation with comoving wave
number $k_0$ enters horizon. Constraints following from both
neutrino and gamma ray experiments are shown. Dashed lines
correspond to the model of critical collapse with the parameters
$\delta_c=0.45, \gamma_c=0.36, k_c=4$, solid lines represent the
results obtained using the standard collapse picture.}
\label{const}
\end{figure}

\section{Results and discussions}

The calculation of the constraints uses two basic observational
facts.

1. The differential energy spectrum of the extragalactic photon
background at energy $E_{\gamma} = 10$MeV is \cite{gam-c}
\begin{equation}
\label{gammaconstr} \sim 10^{-2} GeV^{-1} cm^{-2} s^{-1} sr^{-1}.
\end{equation}

2. According to the data of Super-Kamiokande experiment
\cite{Malek:2002ns}, the electron antineutrino background flux in
space is constrained by the inequality
\begin{equation}
\label{nuconstr} \Phi(E_{\tilde \nu_e}>19.3 GeV) < 1.2 \; cm^{-2}
s^{-1}.
\end{equation}

The resulting constraints are shown on Fig. \ref{const}. The
forbidden values of the most important parameter, $\delta_H(k_0)$,
lie inside the bordered regions drawn on the figure. The value of
the parameter $\Sigma$ characterizing the width of the gaussian
distribution in Eq. (\ref{dHparam}) was fixed throughout all
calculations ($\Sigma = 3$). The constraints are given as a
function of the horizon mass corresponding to the moment of time
when the comoving wave number $k_0$ enters horizon. It follows
from these results that the constraints are stronger in the case
of the standard Carr-Hawking collapse.

The second important result is that the constraints based on
neutrino emission of PBHs are comparable with those following from
photon emission. At the region of small horizon masses (large
$k_0$) where the large red shifts are sufficient, the constraints
from neutrino emission are stronger.

One should add, at the end, one comment. In this work we
calculated the constraints using the rather high value of $T_{RH}$
($=10^{10}$GeV). The corresponding horizon mass at the end of
inflation, $M_i$, is equal to $\sim 10^{11}$g. The interval of
studied here values of the $k_0$ parameter corresponds to PBH
production in super-horizon perturbations. At low $T_{RH}$ the
value of $M_i$ will  be high and the parameter $k_0$ can be chosen
such that black holes will be produced from sub-horizon
perturbations (the possibility of PBH production from such
perturbations is studied in \cite{Lyth:2005ze}). In principle,
constraints on the power spectrum due to PBH evaporations can be
obtained in this case also, using the hypothesis of critical
collapse.

\end{document}